# Exact analysis of potential flow past bodies of irregular shapes

Ankur Jain

Mechanical and Aerospace Engineering Department
University of Texas at Arlington, Arlington, TX, USA

[*] – Corresponding Author: email: jaina@uta.edu; Ph: +1 (817) 272-9338
500 W First St, Rm 211, Arlington, TX, USA 76019

**Abstract:**

Fluid flow past one or more solid bodies is a fundamental problem of much practical importance. Standard solutions of simplified problems involving incompressible inviscid irrotational flow past common geometries such as circular cylinders and airfoils are commonly available. This work presents exact analysis of a potential flow problem involving fluid flow past one or more bodies of irregular shapes. The problem is solved by expressing the shape of each body using Heaviside functions, and writing the potential function as an eigenfunction-based series. Using the properties of Heaviside functions, the series coefficients are determined by deriving a set of linear algebraic equations that govern the coefficients. Benchmarking of the analytical technique against well-known solutions of standard problems is carried out, showing excellent agreement. Good agreement with past work on the specific problem of potential flow past multiple circular cylinders further establishes the accuracy of the analytical technique. Illustrative problems of flow past complicated geometries are solved. Implementation aspects and limitations of the analytical technique are discussed.

**Keywords:** Potential Flow; Irrotational Flow; Analytical Solution; Flow Past a Body.

# 1. Introduction

Fluid flow past one or more solid bodies is a problem of fundamental interest in fluid mechanics [1], with applications in diverse fields such as aerodynamics, hydrodynamics, bioengineering and sports, to name a few. While the general form of this problem is quite complex, involving rotational flow, viscous forces, flow separation and turbulence, several simplifying assumptions are often made to enable engineering analysis. For example, under the assumption of an incompressible inviscid and irrotational fluid, the steady state flow field is given by the gradient of a potential function, which is governed by the Laplace equation, and is considerably easier to analyze than the general Navier-Stokes equations [2]. Despite the underlying simplifying assumptions, analytical solutions of potential flow problems have played a valuable role in several engineering fields, most notably, the development of airfoil theory and wing design [3].

Two-dimensional potential flow problems with relatively simple geometry can be solved analytically using techniques such as conformal transformations, method of images and separation of variables [2]. A variety of realistic flow scenarios can be modeled by linearly combining potential functions of various elementary flows such as sources, sinks and vortex flows. One of the most fundamental analytical results for potential flow past a body is that of flow past a single circular cylinder [2]. Other more complicated geometries can be analyzed, provided an appropriate conformal mapping to transform the given geometry to a single cylinder is available. The Joukowsky transformation [4] has been commonly used for analysis of flow past general airfoils. The Schwarz-Christoffel mapping enables analysis of flow past a simple closed polygon [2]. In addition to these generalized methods, analysis of several specific potential flow problems has also been reported. For example, problems involving flow past two [5] or more [6] cylinders have been solved using conformal mapping. This problem has also been solved using the method of images [7]. Potential flow past a cascade of airfoils has been solved using Riemann–Hilbert analysis [8]. While potential flow theory is applied most commonly to incompressible flows, the use of Prandtl-Glauert and Göthert transformations [9] extend their applicability to subsonic compressible flows as well. A number of numerical techniques are also available for approximate analysis of potential flows. For example, panel methods that use a distribution of singularities on the body surface have been used commonly for aircraft design [10,11]. Numerical analysis based on the method of fundamental solutions [12] and the finite-element method [13] have also been presented.

Despite the considerable literature on potential flow analysis for specific geometries as summarized above, there continues to be interest in the development of general techniques that may address potential flow past a body of arbitrary shape. While, in principle, an appropriate conformal transformation can make it possible to analytically solve a problem with any geometry, the availability of such an analytical conformal transformation is not guaranteed for complicated geometries. Further, the advent of additive manufacturing (3D printing) makes it possible to print parts of nearly arbitrary shape, which further motivates the development of techniques to solve problems of flow past bodies of irregular shapes. Despite the availability of numerical techniques such as panel methods [10.11], the development of analytical techniques to solve the general problem is quite desirable. Analytical techniques provide valuable insights into the fundamental nature of the problem that may not be readily available through numerical simulations. Moreover, compared to numerical techniques, analytical techniques may be more accurate and computationally efficient.

In the context of the discussion above, this work derives an eigenfunction-based series solution for the two-dimensional steady state potential flow of an inviscid irrotational incompressible fluid past one or more bodies of irregular shapes. Benchmarking against well-known solutions for standard problems, as well as comparison with past work for specific problems demonstrates the accuracy of the theoretical technique. The technique is illustrated by solving two problems for which an analytical solution is not available in the past literature.

## 2. Theoretical Analysis

### 2.1. Problem Description

The problem of interest in this work concerns flow of an incompressible inviscid flow past one or more solid bodies of given arbitrary shapes in a two-dimensional domain of size $a$ by $b$, as shown in Figure 1(a). Assuming a total of $\wp \geq 1$ bodies, as shown in detail in Figure 1(b), the shape of the $p^{th}$ body is defined in terms of $x_{Lp}$ and $x_{Up}$, its lower and upper limits in the $x$ direction, respectively, and by functions $y_{Lp}(x)$ and $y_{Up}(x)$ that define the upper and lower bounds of the body in the range $x_{Lp} \leq x \leq x_{Up}$, both as functions of $x$. For example, for a circle of radius $R$ centered at $(x_0, y_0)$, $x_{Lp} = x_0 - R$, $x_{Up} = x_0 + R$, $y_{Lp}(x) = y_0 + \sqrt{R^2 - (x - x_0)^2}$ and $y_{Up}(x) = y_0 - \sqrt{R^2 - (x - x_0)^2}$. A uniform flow with $x$-velocity $U$ enters the domain along the $x = 0$ face. The size of the domain is taken to be sufficiently large compared to the sizes of the bodies, so that the flow returns to its one-dimensional form past the bodies at the $x = a$ face. Additionally, the bottom and top faces at $y = 0$ and $y = b$ are assumed to be impenetrable.

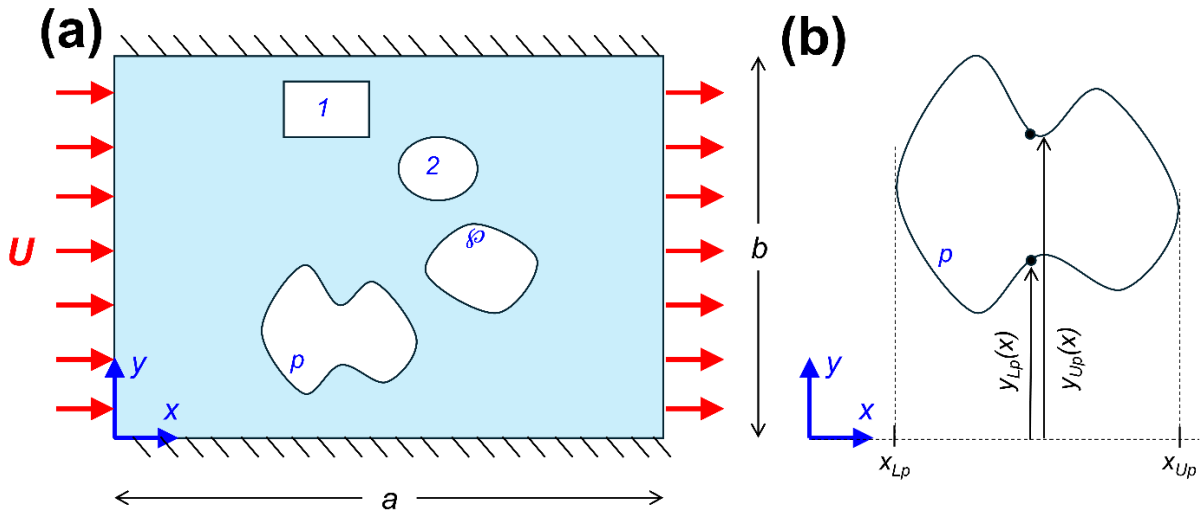

Figure 1. (a) Schematic of the problem of potential flow past a number of solid bodies of irregular shapes considered in this work, (b) Mathematical definition of the $p^{th}$ solid body, in order of its limits $x_{Ap}$ and $x_{Bp}$ in the $x$ direction, and functions $y_{A1}(x)$ and $y_{B1}(x)$ that define its lower and upper boundaries, respectively.

Based on this problem definition, the interest is in determining the velocity field around the bodies. Based on the incompressible, inviscid and two-dimensional nature of flow, it is sufficient to determine the potential function $\phi(x,y)$, the derivatives of which may be used to determine the velocity field. Under the assumptions listed here, the potential function satisfies the Laplace equation. For several special cases of the general problem posed here, such as for a single circular body, the potential function can be determined, for example, using complex variables. Approximate methods to handle other body shapes are also available. The interest here is to develop an exact analytical solution capable of solving this problem for any finite number of bodies, each of which may have arbitrary shape, as defined above.

2.2. Solution Derivation

Based on the problem definition discussed above, one may write the following governing equation for the potential function

$$\frac{\partial}{\partial x}\left(F(x,y)\frac{\partial \phi}{\partial x}\right) + \frac{\partial}{\partial y}\left(F(x,y)\frac{\partial \phi}{\partial y}\right) = 0 \qquad (0 < x < a, 0 < y < b) \qquad (1)$$

where the function $F(x,y)$ must be defined such that $F(x,y) = 1$ in the fluid region and 0 inside any of the solid bodies. The function $F(x,y)$ may be written using Heaviside step functions as follows:

$$F(x,y) = 1 - \sum_{p=1}^{\wp}\left(\mathcal{H}(x - x_{Lp}) - \mathcal{H}(x - x_{Up})\right)\left(\mathcal{H}(y - y_{Lp}(x)) - \mathcal{H}(y - y_{Up}(x))\right) \qquad (2)$$

where $\mathcal{H}(x)$ is the Heaviside step function defined to be 1 if $x > 0$ and 0 otherwise [14]. It may be seen that equation correctly results in $F(x,y)$ becoming 1 everywhere in the fluid region of the domain and zero inside any of the solids. As a result, equation () represents the Laplace equation only within the fluid domain.

The boundary conditions for this problem are

$$\frac{\partial \phi}{\partial x} = U \qquad (x = 0) \qquad (3)$$

$$\phi = 0 \qquad (x = a) \qquad (4)$$

$$\frac{\partial \phi}{\partial y} = 0 \qquad (y = 0, b) \qquad (5)$$

where equation (5) applies a zero potential along the exit face without loss of generality.

The interest is in solving equation () subject to boundary conditions given by equations (3)-(5). In order to do so, one may write the solution for the potential field as follows:

$$\phi(x,y) = U(x - a) + \sum_{n=1}^{\infty}\sum_{m=1}^{\infty} c_{nm} f_n(x) g_m(y) \qquad (6)$$

where the solution may be expressed as an eigenfunction-based series expansion, and the $U(x - a)$ term accounts for the non-homogeneous boundary condition at $x = 0$. The eigenfunctions appearing in the series solution may be chosen to be those of the spatially homogeneous problem corresponding to the problem of interest, i.e., without any solid bodies at all. Based on the boundary conditions, therefore, one may choose $f_n(x) = \cos(\lambda_n x)$ where $\lambda_n = (n - 0.5)\pi/a$ and $g_m(y) = \cos(\mu_m y)$ where $\mu_m = (m - 1)\pi/b$. The eigenfunctions and eigenvalues ensure that boundary conditions given by equations (3)-(5) are satisfied identically. Therefore, it remains to determine the coefficients $c_{nm}$ such that the governing equation for the problem containing the solid bodies is satisfied. Specifically, the solution must account for the shapes of the solid bodies within the domain.

In order to proceed, one may insert the solution given by equation (6) into the governing equation given by equation (1) resulting in

$$\frac{\partial F}{\partial x} \sum_{n=1}^{\infty} \sum_{m=1}^{\infty} c_{nm} f'_n(x) g_m(y) + \frac{\partial F}{\partial y} \sum_{n=1}^{\infty} \sum_{m=1}^{\infty} c_{nm} f_n(x) g'_m(y) \\ - (\lambda_n^2 + \mu_m^2) F(x,y) \sum_{n=1}^{\infty} \sum_{m=1}^{\infty} c_{nm} f_n(x) g_m(y) + U \frac{\partial F}{\partial x} = 0 \quad (7)$$

Further simplification may be obtained based on the following derivatives of $F(x,y)$, obtained from equation (2)

$$\frac{\partial F}{\partial x} = -\sum_{p=1}^{\wp} \Big[ \big(\delta(x - x_{Lp}) - \delta(x - x_{Up})\big)\big(\mathcal{H}(y - y_{Lp}(x)) - \mathcal{H}(y - y_{Up}(x))\big) \\ + \big(\mathcal{H}(x - x_{Lp}) - \mathcal{H}(x - x_{Up})\big)\big(\delta(y - y_{Bp}(x))\big) y'_{Bp}(x) \\ - \delta(y - y_{Ap}(x)) y'_{Ap}(x) \Big] \quad (8)$$

$$\frac{\partial F}{\partial y} = -\sum_{p=1}^{\wp} \big(\mathcal{H}(x - x_{Lp}) - \mathcal{H}(x - x_{Up})\big)\big(\delta(y - y_{Ap}(\xi)) - \delta(y - y_{Bp}(x))\big) \quad (9)$$

Now, equation (7) is multiplied by $f_i(x) g_j(y)$ and integrated over the domain. Substituting equations (8) and (9) in the resulting equation and using orthogonal properties of the eigenfunctions as well as the integral properties of Heaviside and delta functions to simplify, one may obtain a much simplified equation as follows:

$$\sum_{n=1}^{\infty} \sum_{m=1}^{\infty} \big(I_{Anmij} + I_{Bnmij} - (\lambda_n^2 + \mu_m^2) I_{Cnmij}\big) c_{nm} = -I_{Dij} \quad (10)$$

for each $i = 1,2..\infty$ and $j = 1,2..\infty$. Here, the integrals appearing in equation (10) above are given by

$$I_{Anmij} = -\sum_{p=1}^{\wp} \int_{x_{Lp}}^{x_{Up}} f'_n(x)f_i(x)\Big(y'_{Up}(\xi)g_m\big(y_{Up}(x)\big)g_j\big(y_{Up}(x)\big) \\ - y'_{Lp}(x)g_m\big(y_{Lp}(x)\big)g_j\big(y_{Lp}(x)\big)\Big)dx \tag{11}$$

$$I_{Bnmij} = -\sum_{p=1}^{\wp} \int_{x_{Lp}}^{x_{Up}} f_n(x)f_i(x)\Big(g'_m(y_{Lp}(x))g_j\big(y_{Lp}(x)\big) \\ - g'_m(y_{Bp}(x))g_j\big(y_{Bp}(x)\big)\Big)dx \tag{12}$$

$$I_{Cnmij} = \Bigg[\int_0^a f_n(x)f_i(x)dx \int_0^b g_m(y)g_j(y)dy \\ - \sum_{p=1}^{\wp} \int_{x_{Lp}}^{x_{Up}} f_n(x)f_i(x)\Bigg[\int_{y_{Lp}(x)}^{y_{Up}(x)} g_m(y)g_j(y)dy\Bigg]dx\Bigg] \tag{13}$$

$$I_{Dij} = -U\sum_{p=1}^{\wp}\Bigg[\int_{x_{Lp}}^{x_{Up}} f_i(x)\Big(y'_{Lp}(x)g_j\big(y_{Lp}(\xi)\big) - y'_{Up}(x)g_j\big(y_{Up}(\xi)\big)\Big)dx \\ + f_i(x_{Lp})\int_{y_{Lp}(x_{Lp})}^{y_{Up}(x_{Lp})} g_j(y)dy - f_i(x_{Up})\int_{y_{Lp}(x_{Up})}^{y_{Up}(x_{Up})} g_j(y)dy\Bigg] \tag{14}$$

Assuming truncation of the infinite series given by equation (10) up to $N$ terms in both $x$ and $y$ directions, equation (10) represents $N^2$ linear algebraic equations in $N^2$ unknowns – $c_{ij}$ for $i = 1,2,..\infty$ and $j = 1,2,..\infty$. These equations can be solved easily. In matrix notation, $\boldsymbol{c}$ denotes the vector comprising the unknown coefficients, then equation (10) may be written as

$$\boldsymbol{Ac} = \boldsymbol{b} \tag{15}$$

so that an analytical solution for the coefficient vector is given simply by $\boldsymbol{c} = \boldsymbol{A}^{-1}\boldsymbol{b}$. Once solved, the potential field is given by equation (), and the velocity field may be obtained simply by differentiation of the potential field, i.e.,

$$u_x(x,y) = U + \sum_{n=1}^{N}\sum_{m=1}^{N} c_{nm} f'_n(x) g_m(y) \tag{16}$$

$$u_y(x,y) = \sum_{n=1}^{N}\sum_{m=1}^{N} c_{nm} f_n(x) g'_m(y) \qquad (17)$$

This completes the solution of the problem of potential flow around a number of bodies of arbitrary shape. The solution derived here is exact and places no restriction on the number of bodies, their locations, sizes and shapes, as long as the shape of each body is expressible in the form of functions $y_{Ap}(x)$ and $y_{Bp}(x)$. A detailed discussion of implementation aspects, including the handling of different types of body shapes is discussed in section 3.4.

## 3. Results and Discussion

### 3.1. Comparison with Standard Solutions

Standard analytical techniques based on complex analysis and conformal mapping are available for solving potential flow problems for relatively simple geometries, such as flow over a single elliptical/cylindrical cylinder. Specifically, for an elliptical cylinder of axis lengths $2R_x$ and $2R_y$ in the $x$ and $y$ directions, respectively, the potential function is given by $\phi(x,y) = \text{Re}(\hat{z} + \hat{R}^2/\hat{z})$ [2], where $\hat{z} = \frac{1}{2}(z + \sqrt{z^2 - 4c^2})$ and $z = x + iy$ is the complex coordinate with respect to the center of the ellipse. Additionally, $\hat{R} = \frac{1}{2}(R_x + R_y)$ and $c = \frac{1}{2}(z + \sqrt{R_x^2 - R_y^2})$. For the specific case of a circle $R_x = R_y = R$), the potential function is given by $\phi(x,y) = \text{Re}(z + \hat{R}^2/z)$ [2].

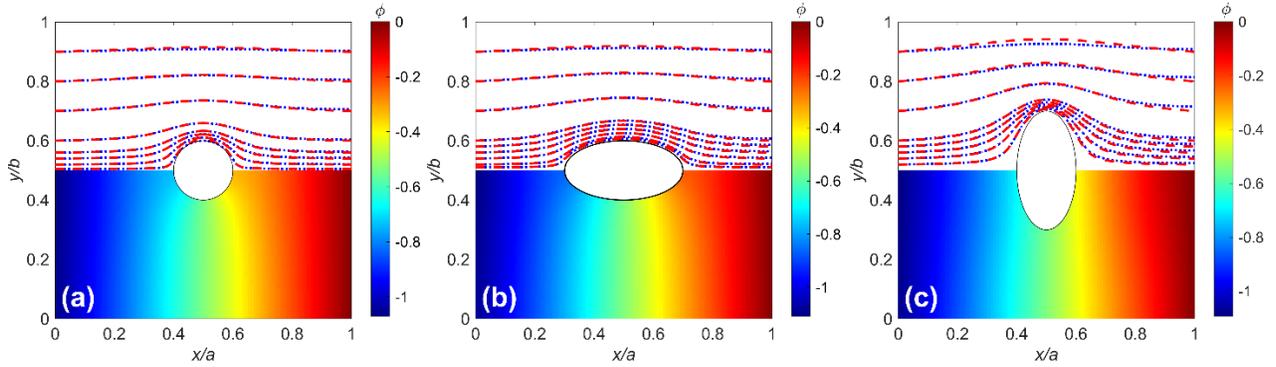

Figure 2. Potential function and streamlines for the standard problems of flow over a cylinder of various cross-sections: (a) circle, (b) horizontal ellipse and (c) vertical ellipse. Multiple streamlines from the present work (blue dotted curves) and the standard solution (red dashed curves) are shown in the top half. In each case, the symmetric potential function distribution is shown in the bottom half.

In order to benchmark the theoretical technique developed in this work, special cases with a single ellipse or circle are solved and compared with solutions above. Figures 2(a)-(c) present the flow streamlines computed from the present method as well as standard solutions for a circle ($R = 0.2a$), a flat ellipse ($R_x = 0.4a, R_y = 0.2a$), respectively, with $a = b$. Due to symmetry, the top half of each plot presents a comparison of the streamlines, while the bottom face presents a colormap of the potential field, found to be practically identical for the present work and the well-known solutions, with the worst-case deviation being less than 1%. Since the solution is in the form of an infinite series, in principle, the deviation can be further reduced by considering a greater number of terms in the series solution. Figure 2 verifies that the solution of the general problem derived here correctly reduces to the well-known solutions of flow past a single circular/elliptical cylinder under special conditions.

4.2. Comparison with Past Work

While the previous section benchmarks the present work against well-known solutions for simplified special cases of the general problem considered here, it is also of interest to compare the present work with available literature on computing the potential field for other more complicated problems. For example, Crowdy [6] has presented an analytical solution based on complex analysis for flow past a number of circular bodies of arbitrary sizes and locations. The problem solved by Crowdy [6] is a special case of the general problem solved in this work. Therefore, it is pertinent to compare the two for the conditions assumed by Crowdy. For this purpose, two specific problems illustrated by Crowdy are considered – two aligned circles of equal size and three unaligned circles of different sizes. These problems are solved using the general analytical technique developed here and results are compared with Crowdy in Figures 3(a) and 3(b). The flow streamlines computed using the present work, shown as dashed lines, are found to be in excellent agreement with Crowdy's results, shown as circles. Figures 3(a) and 3(b) also present colorplots of the potential field computed using the present work, shown only in one half for Figure 3(a) due to symmetry. The good agreement between the present work and a representative past result for a special case of the general problem considered here provides addition confidence in the theoretical technique developed here.

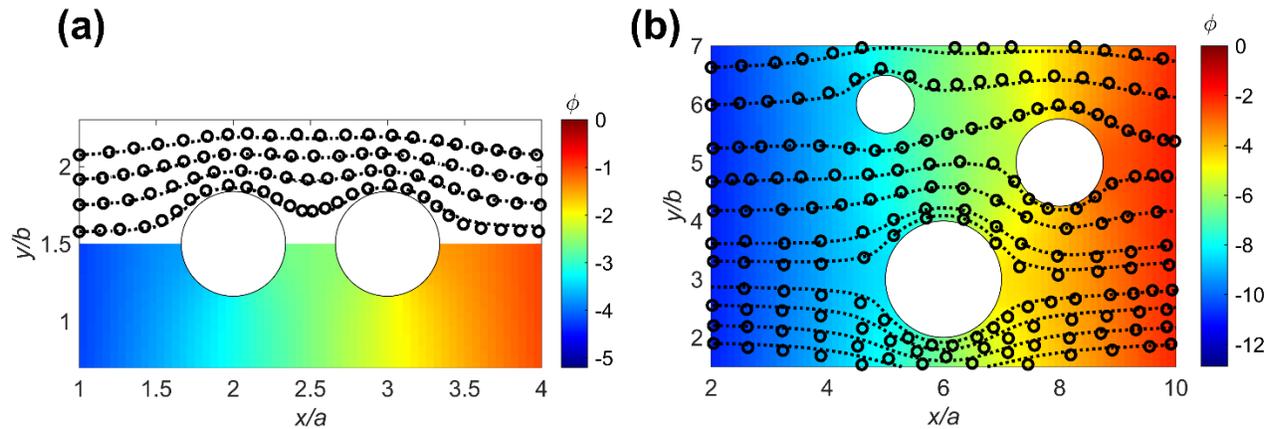

Figure 3. Comparison with conformal transformation based analysis of multiple cylinders [6]: (a) Two horizontally aligned cylinders of equal size, (b) Three unaligned unequally sized cylinders. Multiple streamlines from the present work (dashed curves) and past work based on conformal transformation (circles) are shown. Colorplot represents the potential function, shown only in one half for (a) based on symmetry.

## 4.3. Application of the Technique for Geometrically Complicated Problems

While the previous two sub-sections presented results for relatively simpler geometries that represent special cases of the general problem solved here, it is helpful to illustrate the capabilities of the theoretical technique developed here by solving problems with more complicated geometries. To do so, the potential field is computed for a problem with flow past a cascade of three NACA 5317 airfoils. Results are presented in Figure 4(a), showing flow streamlines over a colorplot of the potential field. Figure 4(a) illustrates the capability of the theoretical technique to account for complicated geometries and predict the key features of potential flow, including bending of flow streamlines around bodies and re-straightening of the flow field past the bodies.

As another illustration, flow past a novel heart-shaped body is considered next. The equation of the heart-shaped body is given by

$$y_{L1}(x) = 0.5 + \frac{\sqrt{|5x - 2.5|} - \sqrt{1 - (5x - 2.5)^2}}{10} \tag{18}$$

$$y_{U1}(x) = 0.5 + \frac{\sqrt{|5x - 2.5|} + \sqrt{1 - (5x - 2.5)^2}}{10} \tag{19}$$

Based on this, Figure 4(b) presents a colorplot of the resulting potential field and flow streamlines past the heart-shaped body. As expected, the flow bends around the heart shape and is largely unaffected sufficiently far from the body.

While a cascade of airfoils and heart-shaped body are used as examples, Figures 4(a) and 4(b) illustrate the capability of the analytical model to compute the potential flow field past bodies of complicated geometries that may be difficult/impossible to address using previously available analytical techniques such as complex analysis.

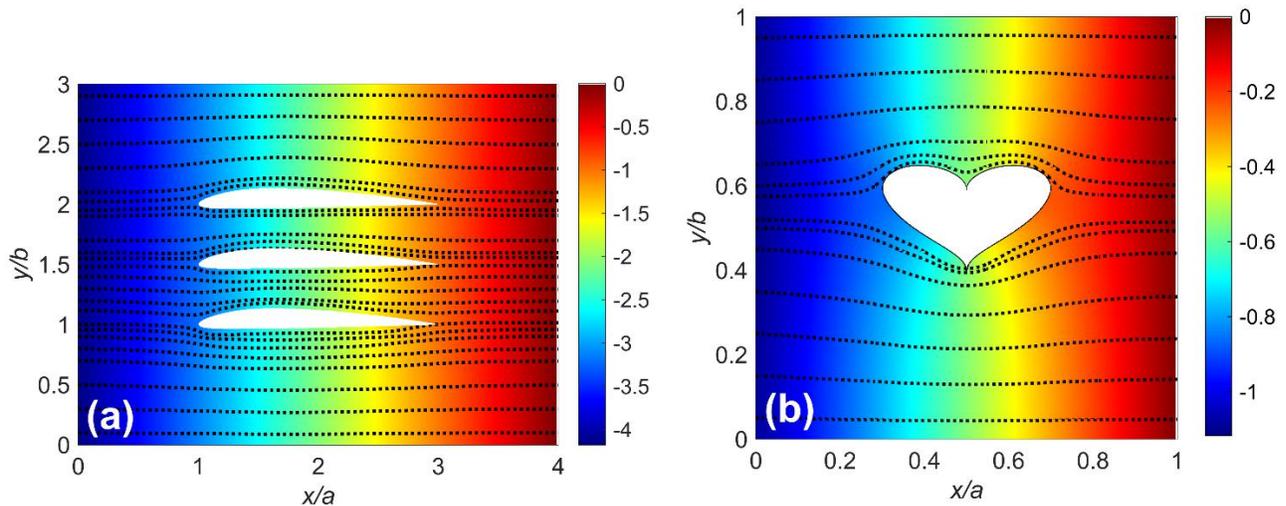

Figure 4. Illustration of the capability to compute potential flow over complicated geometries: (a) a cascade of three NACA 5317 airfoils, (b) a heart-shaped body.

## 3.4. Discussion

Within the limitations of potential flow assumptions, the results derived in section 2 are exact. However, the results do not account for a moving, rotating or deforming solid body. The convergence of the series solution derived here likely depends on the shapes of the bodies, typically requiring greater number of terms to model more complicated shapes, particularly with sharp edges. The order of the matrix needed to be inverted is $N^2$ and the number of integrals to be computed scales as $N^4$. Therefore, the number of terms considered in the series solution strongly affects computational cost. To illustrate the impact of $N$ on accuracy, a representative problem of flow over a single circular cylinder is considered. The potential function is determined with different values of $N$. Figure 5(a) plots two flow streamlines close to the cylinder for different values of $N$. This plot shows that the nature of streamlines changes with increasing $N$ and converges around $N = 14$. This is further illustrated in a plot of the potential function at a specific point as a function of $N$, presented in Figure 5(b). This plot shows convergence with increasing number of terms. It is likely that, due to the representation of the potential function with periodic functions, a greater number of terms may be needed for more complex shapes, especially those with non-smooth boundaries such as rectangles. A formal convergence analysis for the results derived here may be considerably mathematically challenging due to the general nature of matrix $\boldsymbol{A}$ and the complicated expressions for various integrals that comprise its elements.

The integrals given by equations (11)-(14) may be determined explicitly for relatively simple geometries, but for more practical geometries, a numerical computation may be needed. Symmetry within several integrals defined in these equations, such as $I_{Anmij} = I_{Aimnj}$ and $I_{Cnmij} = I_{Cimnj} = I_{Cnjim}$ may be useful for reducing the cost of computing the potential function using this method.

The treatment of certain body shapes is noteworthy. $I_{Anmij}$ and $I_{Dij}$ require integrals involving derivatives of functions $y_{Lp}(x)$ and $y_{Up}(x)$ that govern the shape of the body. In case these derivatives becomes unbounded at $x = x_{Lp}$ or $x = x_{Up}$, for example, in case of a circle, one may still evaluate the resulting improper integral of type 2 by replacing $x_{Lp}$ by $x_{Lp} + \Delta x$ and/or $x_{Up}$ by $x_{Up} - \Delta x$ and considering the limit in which $\Delta x$ goes to zero. In a practical numerical implementation of this technique, one may move the shift the limit of integration by a very small quantity in order to avoid the unbounded derivative. If the derivative becomes unbounded between $x = x_{Lp}$ and $x = x_{Up}$, for example, due to a sharp edge, then one may split up the integral into two or more parts and evaluate each based on the strategy above.

The second terms on the right hand sides of equations (11) and (14) account for vertical boundaries in shapes such as rectangles. If the shape of body of interest is continuous at the $x$ ends, i.e., if $y_{Lp}(x_{Lp}) = y_{Up}(x_{Lp})$ and $y_{Lp}(x_{Up}) = y_{Up}(x_{Up})$, then these terms become zero, leading to computational simplification.

Finally, in case of 3D printed bodies, the functions $y_{Lp}(x)$ and $y_{Up}(x)$ may be defined in the form of a collection of points on the surface. In such a case, the derivatives of these functions appearing in equations (11) and (14) may need to be computed using finite differencing, and the equations may need to evaluated numerically.

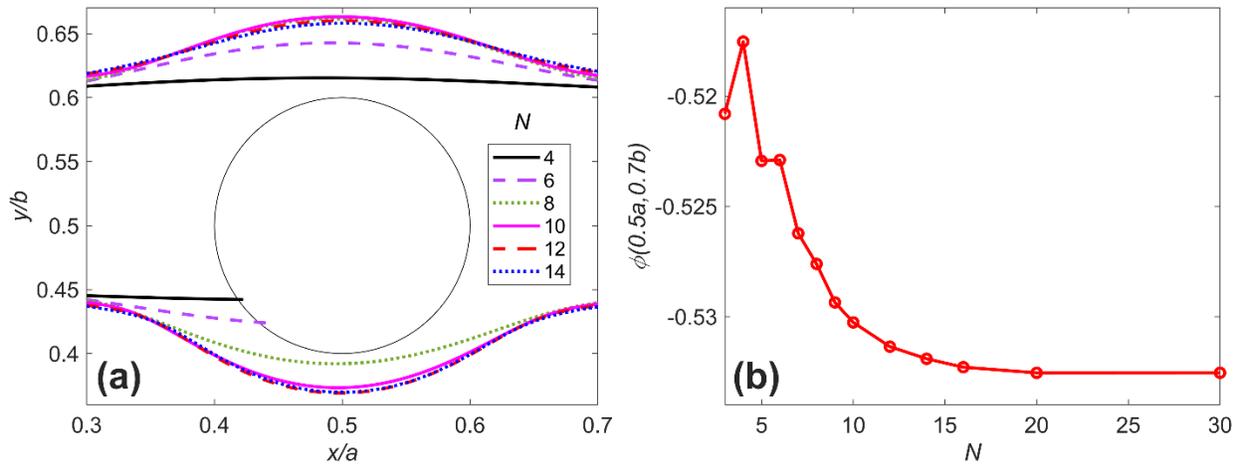

Figure 5. Effect of number of terms in the series solution for the problem of flow past a circular cylinder of radius $0.1a$ with $a = b$: (a) two streamlines for different values of the number of terms in the series solution, (b) potential function at $(0.5a, 0.7b)$ as a function of number of terms in the series solution.

A particular advantage of the technique developed here is that the eigenvalues of the problem can be easily determined, independent of the geometry of the bodies. Also, the definition of $F(x,y)$ being 1 in the fluid domain and 0 inside the solids ensures that the key interface condition of zero normal velocity at the surface of the solid bodies is satisfied regardless of how complex the shape of the solid surface may be, even though it is not enforced explicitly. This is because, regardless of the nature of $F(x,y)$ in equation (1), the Laplace equation being solved here admits a continuous and differentiable solution, comprising periodic functions as shown in equation (2). Therefore, the potential function as well as the velocity field are expected to be continuous throughout, including at the surfaces of the solid bodies. Now, $F(x,y)$ in the current problem plays the same role as thermal conductivity in an equivalent thermal transport problem [15]. A zero value of $F(x,y)$ inside each solid bodies, enforced by equation (2) is analogous to a zero thermal conductivity material in the thermal transport problem, which does not allow any flow of heat normal to the body surface. By analogy, thus, the potential function determined from the technique developed here automatically ensures zero normal velocity at the surfaces of the bodies.

A clear limitation of this work is that each body must be two-dimensional, and not be moving or rotating. Shapes such as a thin plate can not be described by equation (2) and thus can not be analyzed using this technique. Moreover, fractal shapes require complicated definitions of $y_{Lp}(x)$ and $y_{Up}(x)$, potentially leading to computational difficulties.

## 5. Conclusions

Potential flow past bodies of general irregular shapes can be analyzed using the technique developed here. This represents a significant generalization of past work that is limited to specific body shapes. Establishing the convergence of the series solution is recommended for future work, as is the implementation of this technique for specific problems of engineering interest.

**References**

[1] Batchelor, G.K. (2000) *An Introduction to Fluid Dynamics*, Second Ed., Cambridge Mathematical Library.

[2] Marshall, J.S. (2001) *Inviscid Incompressible Flow*, First Ed., Wiley-Interscience.

[3] Thwaites, B. (1987) *Incompressible Aerodynamics - An account of the theory and observation of the steady flow of incompressible fluid past aerofoils, wings and other bodies*, First Ed., Dover Publications.

[4] Joukowsky, N.E. (1910) Über die Konturen der Tragflächen der Drachenflieger, *Zeitschrift für Flugtechnik und Motorluftschiffahrt*, **1**, pp. 281-284.

[5] Burton, D.A., Gratus, J. & Tucker, R.W. (2004) Hydrodynamic forces on two moving discs, *Theoret. Appl. Mech.*, **31**, pp. 153-187.

[6] Crowdy, D.G. (2006) Analytical solutions for uniform potential flow past multiple cylinders, *Europ. J. Mech. B/Fluids*, **25**, pp. 459-470.

[7] Dalton, C. & Gelfinstine, R.A. (1971) Potential flow past a group of circular cylinders, *Trans. ASME J. Basic Eng.*, **93**, pp. 636-642.

[8] Baddoo, P.J. & Ayton, L.J. (2018) Potential flow through a cascade of aerofoils: direct and inverse problems, *Proc. Roy. Soc. A*, **474**, pp. 2018:0065.

[9] Shapiro, A.H. (1953) *The dynamics and thermodynamics of compressible fluid flow*, First Ed., Wiley.

[10] Hess, J.L. (1975) Review of integral-equation techniques for solving potential-flow problems with emphasis on the surface-source method, *Computer Meth. Appl. Mech. Eng.*, **5**, pp. 145-196.

[11] Hess, J.L. & Smith, A. (1967) Calculation of potential flow about arbitrary bodies, *Prog. Aero. Sci.*, **8**, pp. 1-138.

[12] Johnston, R.L. & Fairweather, G. (1984) The method of fundamental solutions for problems in potential flow, *Appl. Math. Modeling*, **8**, pp. 264-270.

[13] Argyris, J.H. & Mareczek, G. (1972) Potential flow analysis by finite elements, *Ingenieur-Archiv.*, **42**, pp. 1-25.

[14] Abramowitz, M. & Stegun, I. (1970), *Handbook of Mathematical Functions*, Ninth Ed., Dover Publications.

[15] Jain, A. & Krishnan, G. (2025) Exact analytical solution for thermal conduction in a Cartesian body with heat-generating regions of arbitrary shapes and thermal properties, *Int. J. Heat Mass Transf.*, **225**, pp. 459-470.